\documentclass[journal]{IEEEtran}
\ifCLASSINFOpdf
\else
\fi
\usepackage{graphicx}  
\usepackage{bm}        
\usepackage{amssymb}   
\usepackage{amsmath}
\usepackage{float}
\usepackage{color}
\usepackage{physics}

\begin{document}
%
\title{Fundamental Limitations of Rayleigh Backscattering Noise on Fiber-Based Multiple-Access Optical Frequency Transfer}
%
%
%

\author{Liang Hu, Xueyang Tian, Guiling Wu,~\IEEEmembership{Member,~IEEE,} Jianguo Shen, and Jianping Chen
\thanks{Manuscript received xxx xxx, xxx; revised xxx xxx, xxx. This work was supported in part by the National Natural Science Foundation of China (NSFC) (61627871, 61535006, 61905143), in part by science and technology project of State Grid Corporation of China (No. SGSHJX00KXJS1901531) and in part by Zhejiang provincial Nature Science Foundation of China (LY17F050003). (\textit{Corresponding author: Liang Hu and Guiling Wu})}
\thanks{L. Hu, X. Tian, G. Wu, J. Chen are with the State Key Laboratory of Advanced Optical Communication Systems and Networks, Department of Electronic Engineering, Shanghai Jiao Tong University, Shanghai 200240, China, and also with  Shanghai Institute for Advanced Communication and Data Science, Shanghai Jiao Tong University, Shanghai 200240, China (e-mail: liang.hu@sjtu.edu.cn; txy0220@sjtu.edu.cn; wuguiling@sjtu.edu.cn; jpchen62@sjtu.edu.cn).}
\thanks{J. Shen is with the College of Physics and electronic information Engineering, Zhejiang Normal University, Jinhua, 321004, China (e-mail: shenjianguo981@163.com).}
}

%
%

\markboth{JOURNAL OF LIGHTWAVE TECHNOLOGY,~Vol.~xxx, No.~xxx, March~2020}%
{Shell \MakeLowercase{\textit{et al.}}: Bare Demo of IEEEtran.cls for IEEE Journals}
%



\maketitle

\begin{abstract}
Ultimate limits of performance for single-source-bidirectional architecture e.g., optical frequency transfer are set by delay fluctuations within the host material, e.g., optical fiber or atmospheric channel, which is dynamically induced. The most common manifestation of such phase disorder in the optical fiber is backscattering lights coming from Rayleigh backscattering and fresnel reflection, which is observed in nearly all single-source-bidirectional architectures and can lead to both irreversible coherence losses as well as undesirable interference coupling. While it has been shown that backscattering induced phase noise can be suppressed by adopting acoustic-optic-modulators (AOMs) at the local and remote sites to break the frequency symmetry in both directions. However, this issue can not be avoided for conventional fiber-optic multiple-access coherent optical phase dissemination in which the interference of the signal light with the Rayleigh backscattered light will probably destroy the coherence of the stabilized optical signal.  We suppress the backscattering effect by locally breaking the frequency symmetry at the extraction point  by inserting an additional AOM. Here, we theoretically analyze and experimentally demonstrate an add-drop one more AOM approach for suppressing the Rayleigh backscattering within the fiber link. Near-complete suppression of backscattering noise is experimentally confirmed through the measurement—the elimination of a common interference term of the signal light and the Rayleigh backscattered light. The results demonstrate that the Rayleigh backscattering light has a limited effect compared to the residual delay-limited fiber phase noise on the system's performance. Our results also provide new evidence that it is possible to largely suppress Rayleigh and other backscattering noise within a long optical fiber link, where the accumulated phase noise could be large, by using frequency symmetry breaking at each access node to achieve robust multiple-access coherent optical phase propagation in spite of scatters or defects.


\end{abstract}

\begin{IEEEkeywords}
Optical clock, Rayleigh backscattering, multiple-access optical frequency transfer, metrology.
\end{IEEEkeywords}

%
\IEEEpeerreviewmaketitle

\section{Introduction}

\IEEEPARstart{A}{tomic} clocks, in particular optical clocks, are advancing the frontier of measurement science, enabling searches for dark matter and physics beyond the Standard Model \cite{derevianko2014hunting, van2015search, arvanitaki2015searching, hu2017atom}, as well as providing innovative quantum technologies for other branches of science \cite{ludlow2015optical, safronova2018search, poli2014optical}. Over the past few decades, several frequency transfer schemes have been developed \cite{allan1980accurate, ma1994delivering} and  optical fiber has been recognized as a more promising solution than satellite-based techniques \cite{ma1994delivering}.  Currently, researchers are mainly focusing on long-distance connections between just two locations connected by an optical fiber to correct phase perturbations at the local site \cite{ma1994delivering, Riehle:2017aa} used  for atomic clock comparisons \cite{lisdat2016clock, delva2017test}, enabling transfer distance to up to a few hundred of kilometers \cite{newbury2007coherent, grosche2009optical, foreman2007coherent, jiang2008long, yamaguchi2011direct, calosso2014frequency}. To satisfy the requirements of multiple-point optical frequency transfer applications, such as precise measurements of general relativity and temporal variation of the fundamental constant \cite{parker2018measurement}, multiple-access optical frequency transfer was proposed and demonstrated to extract the ultrastable signal for multiple users along the main link \cite{bercy2014line, grosche2014eavesdropping, krehlik2013multipoint, grosche2010method}, allowing extracting  the fiber anywhere and deriving an optical reference frequency with the same precision as that achieved at the end point. 

In these pioneering works, they provided a fixed angular frequency shift at the remote site with an acousto-optic modulator (AOM) to distinguish light reflected at the remote end from stray reflections along the link, such as Rayleigh backscattering and fresnel reflections \cite{ma1994delivering}. This principle is also adaptable for ultrastable optical frequency dissemination schemes on a star topology optical fiber network \cite{schediwy2013high, wu2016coherence, hu2020multi}. Using this method, a highly synchronized optical signal itself can be recovered at any remote site by actively compensating the phase noise of each fiber link at each user end \cite{schediwy2013high, wu2016coherence, hu2020multi}. However, this avoiding backscattering technique cannot be directly applied to the access nodes where they are located between the local and remote AOMs by extracting bidirectional transferred optical signals. Consequently, noise principally due to beating between the extracted signals and the backscattered signals can not be avoided.

\begin{figure*}[htbp]
\centering
\includegraphics[width=\linewidth]{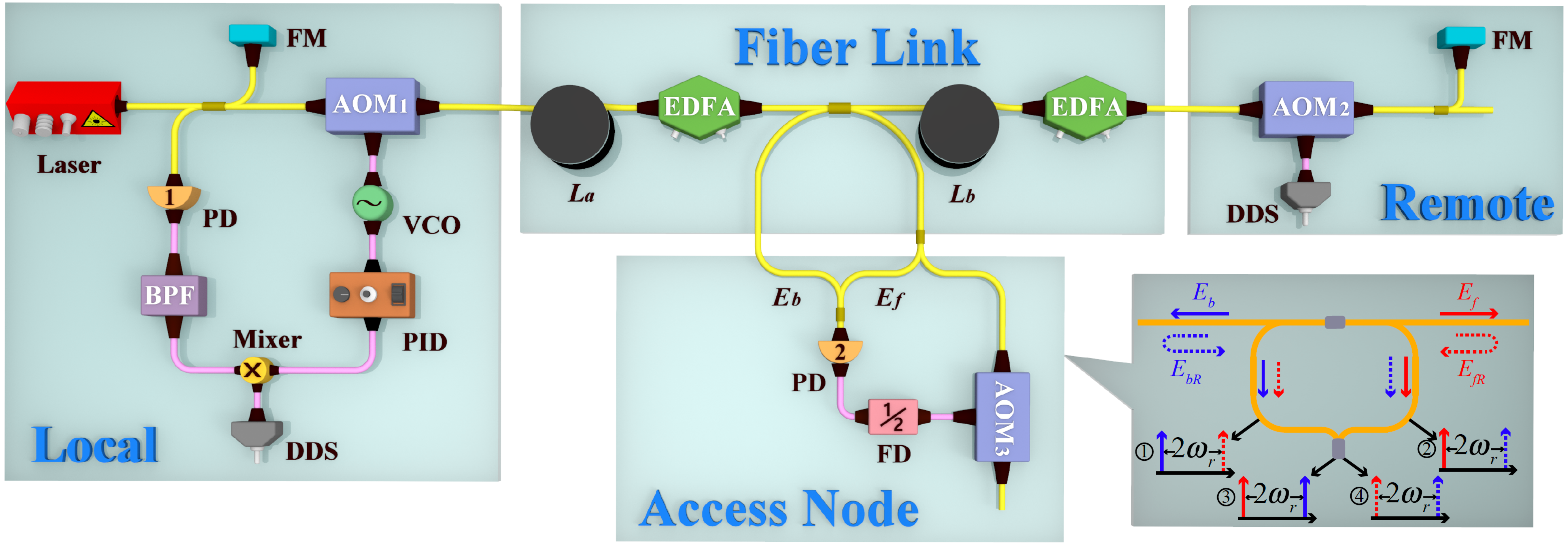}
\caption{Schematic diagram of multi-access optical frequency transfer.  A stabilized optical signal can be obtained by extracting the froward and backward signals from the actively stabilized main fiber link.  The yellow and pink lines represent optical and electrical signals, respectively. The lower right inset illustrates the effect of the Rayleigh scattering  on the PD2. The solid and dashed lines represent the extracted and backscattered signals, respectively, and the red and blue curves represent the forward and backward signals, respectively.  \textcircled{\raisebox{-0.9pt}{1}}-\textcircled{\raisebox{-0.9pt}{4}} terms represent the interference of the extracted light with the backscattered  light. AOM: acousto-optic modulator, FM: Faraday mirror, DDS: direct-digital synthesizer, EDFA: erbium-doped fiber amplifier, FD: frequency divider, PID: proportional–integral–derivative controller, PD: photo-detector, VCO: voltage-controlled oscillator.}
\label{fig1}
\end{figure*}

 In such systems, the noise that is induced by the backscattering light may deteriorate the performance of bidirectional transfer systems \cite{staubli1991crosstalk} since the backscattering adds significant noise. The main factors that lead to backscattered propagating waves in optical fibers are fresnel reflection, Rayleigh and Brillouin scattering \cite{xi2007optical, schroeder1973rayleigh}. When the incident optical power in the fiber is weaker than a few dBm, the dominant effect is the Rayleigh scattering caused by inhomogeneities in the refractive index \cite{laberge1973equilibrium, schroeder1973rayleigh, watanabe2003fictive, le2008density}.  While optical connectors with oblique end faces and fusion splices can have return losses of lager than 55 dB, the Rayleigh backscattering of about $-32\sim-33$ dB of the launched power can not be avoided. Severe problems in optical communication due to noise induced by the Rayleigh backscattering have been widely investigated \cite{burns1983rayleigh, sliwczynski2012bidirectional, lee2005bidirectional}. For example, the effect of the Rayleigh backscattering in fiber gyroscopes and performance degradations in bidirectional optical communication systems caused by the backscattered light have been investigated \cite{burns1983rayleigh}. Furthermore, the Rayleigh backscattering may limit the use of optical amplifiers in fiber systems \cite{sliwczynski2012bidirectional, lee2005bidirectional}. Therefore, multiple reflections along the fiber path give rise to the mixture of the noise terms and the desired signal, which could  cause performance degradations in multi-access optical frequency transfer systems, and should be thoughtfully investigated.

In this article, we first present a comprehensive model to describe the phase noise that is induced by the Rayleigh backscattering in a multi-access optical frequency  system. Our comprehensive model takes into account the main factors required to accurately model the Rayleigh backscattering in optical fibers for multi-access optical frequency transfer. In particular, we include in the model the phase noise due to thermal fluctuations, which is longer than the round-trip propagation time of the light in the fiber. According to the numerical simulation results, we found that the Rayleigh backscattering phase noise because of large accumulated thermal fluctuations of waves propagating in the fiber does not significantly affect the access-node performance in conventional multiple-access optical frequency transfer without multiple amplifiers in the fiber link. Therefore, we infer that the Rayleigh backscattering induced phase noise does not have much impact on the multiple-access optical frequency transfer. The theoretical results were compared to experimental results that were obtained for multi-access optical frequency transfer over a 100 km fiber link where the access node was located after a 50 km fiber link. The experimental results confirm that the backscattering noise dominated by the Rayleigh backscattering has a neglected effect compared to the delay-limited fiber phase noise on the performance of multi-access optical frequency transfer at the precision of our experiment.

The article is organized as follows. We illustrate the concept of multi-access optical frequency transfer with the Rayleigh backscattering effect in Sec. II. In Sec. III the sensitivity degradations at the access node due to the Rayleigh backscattering noise are calculated. A description of the performed measurements and a comparison of the theoretical calculations with the experimental results are given in Section IV. Finally, Section V and VI contain the discussion and conclusion of this study, respectively.

\begin{figure*}[htbp]
\centering
\includegraphics[width=\linewidth]{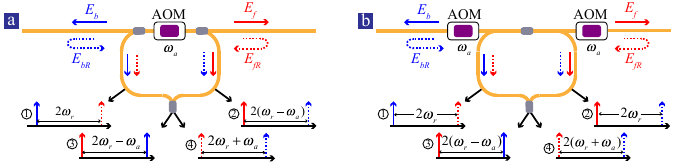}
\caption{Improved extraction setups by inserting one AOM or two AOMs at the extraction point. (a) One AOM: Replacing the $2\times2$ coupler with an AOM (downshifted mode, $-1$ order) fed by an RF signal with the angular frequency of $\omega_a$ and two $1\times2$ couplers at the extraction point.  (b) Two AOMs: Inserting two opposite frequency-shifted AOMs (upshifted mode, $+1$ order and downshifted mode, $-1$ order, respectively) at both sides of the $2\times2$ coupler. The latter scheme guarantees that no frequency modifications will be implemented on the main fiber link. Here we assume $\omega_r>\omega_l>\omega_a$. }
\label{fig2}
\end{figure*}

\section{Concept of multiple-access optical frequency dissemination}
\subsection{Basic principle}

Figure \ref{fig1} illustrates a schematic diagram of multiple-access optical frequency dissemination that has been previously reported \cite{grosche2014eavesdropping, bercy2014line, bercy2016ultrastable, bai2013fiber}.  Here we briefly describe our configuration. An optical-frequency $\nu$ from the signal source  is downshifted by an angular frequency $\omega_l$ with an AOM denoted as AOM$_1$. The light then feeds into the fiber. At the remote site, it is upshifted by an angular frequency $\omega_r$ with AOM$_2$, and part of it is sent back with a Faraday mirror (FM). The round-trip signal is mixed with the input ultra-stable laser with the assistance of an optical coupler and another FM onto a photo-detector (PD1). The beat-note frequency is twice the sum of the AOMs frequencies, which includes the round-trip fiber phase noise, $2\phi_p$. Here we assume that the phase noise introduced by the fiber link in each direction is $\phi_p$. After downconversion to zero frequency with a standard radio frequency (RF) signal, the phase correction $\phi_c=-\phi_p$ is applied to the RF port of AOM$_1$. Eventually at the remote site, the phase fluctuations are $\phi_c+\phi_p$ and thus cancelled. 

For the access node at a distance $L_a$ from the input end and $L_b$ from the output end, a $2\times2$ optical coupler can be adopted to extract both the forward and backward signals from the fiber. The forward signal has a frequency of $\nu-\omega_{l}$, with a phase fluctuation of $-\phi_p+\phi_a=-\phi_b$, demonstrating that the main link compensation introduces overcorrection of the fiber noise on the extracted forward signal. In order to compensate for this overcorrection, we detect the beat-note between the two extracted signals from both sides. The backward signal has a frequency of  $\nu-\omega_{l}+2\omega_{r}$,  exhibiting a phase fluctuation $-\phi_p+\phi_a+2\phi_b=\phi_b$. The beat-note frequency between the extracted forward and backward signal is thus $2\omega_r$ and exhibits the phase fluctuation $2\phi_b$. The beat-note is divided by a factor of 2 and then filtered to drive an AOM (AOM$_3$) in order to correct the phase fluctuations of the extracted forward signal $\nu-\omega_l$. The frequency of the extracted forward signal, after passing through AOM$_3$, is thus upshifted to $\nu$ and its phase fluctuation cancelled. Similar compensation can be obtained on the extracted backward  signal with a negative optical frequency shifter.

The above description has ignored the effect of the Rayleigh noise on the detected RF signal. The Rayleigh noise backscattered from the forward and backward optical signals will, respectively, be superimposed onto the extracted backward and forward signals as illustrated in the lower right inset of Fig. \ref{fig1}. We can see that the beat-note between the extracted signals and the backscattered  signals (``\textcircled{\raisebox{-0.9pt}{1}}'' and ``\textcircled{\raisebox{-0.9pt}{2}}'') will also produce the same detected RF frequency of $2\omega_r$ as the  desired one (``\textcircled{\raisebox{-0.9pt}{3}}''). Here, we use a scalar field in the following derivations for clarity.  The total electrical field $E(t)$ at the PD2 is given by,
\begin{equation}
E(t)=E_b(t)+E_{f}(t)+E_{bR}(t)+E_{fR}(t),
\label{eq1}
\end{equation}
where $E_b(t)$, $E_{f}(t)$, $E_{bR}(t)$ and $E_{fR}(t)$ are the extracted backward signal , the extracted forward signal, the backscattered backward signal and  the backscattered forward signal. An electrical signal recovered by the PD2 can be expressed as,
\begin{equation}
\begin{split}
{E_{\text{RF}}}(t)&\propto |E_b(t)E^*_{fR}(t)|+|E_f(t)E^*_{bR}(t)|\\
&+|E_b(t)E^*_f(t)| +|E_{fR}(t)E^*_{bR}(t)|,
\end{split}
\label{eq2}
\end{equation}
where we ignore the conversion factor related to the photo-detection process. As shown in the lower right inset  of Fig. \ref{fig1}, the first term  is the beat-note between $E_b(t)$ and $E_{fR}(t)$ (``\textcircled{\raisebox{-0.9pt}{1}}''), the second term is the beat-note between $E_f(t)$ and $E_{bR}(t)$ (``\textcircled{\raisebox{-0.9pt}{2}}''), the third term is the beat-note between $E_b(t)$ and $E_{f}(t)$ (``\textcircled{\raisebox{-0.9pt}{3}}''), and the last term is the beat-note between $E_{bR}(t)$ and $E_{fR}(t)$ (``\textcircled{\raisebox{-0.9pt}{4}}'').  The beat signals between $E_b(t)$ and $E_{bR}(t)$ and  between $E_f(t)$ and $E_{fR}(t)$ convert to zero frequency and can removed by a bandpass filter. As the power of the fourth term is much less than the first three terms, its effect can be ignored here. We divide the angular frequency  of  ${E_{\text{RF}}}(t)$ by a factor of 2 and feed it onto the RF port of the AOM$_3$. Thus the phase corrected optical signal $E_E$ at the access node output can be expressed as,
\begin{equation}
\begin{split}
{E_{\text{E}}}(t)&\propto |E_{f}E_{b}|\cos(\nu t)+|E_{f}E_{bR}|\cos(\nu t+{\phi_{bR}}/{2}+\phi_1)\\
&+|E_{b}E_{fR}|\cos(\nu t-{\phi_{fR}}/{2}+\phi_2)\propto \cos(\nu t+\phi_c)
\end{split}
\label{eq3}
\end{equation}
where the  time independent electrical field symbol represents its amplitude, $\phi_c$ is expressed as an arctan function in Eq. \ref{eq4} and we have ignored the nonlinear effect and gain flatness of the amplifiers and the photo-detectors. It is important to stress that the initial phase offsets between the extracted optical signals and the backscattered optical signals ($\phi_1$ and $\phi_2$) have to be included into the numerical simulation.


\begin{figure*}[htbp]
\begin{equation}
\phi_c =\text{arctan}\frac{|E_{f}E_{bR}|\sin({\phi_{bR}}/{2}+\phi_1)+|E_{b}E_{fR}|\sin({-\phi_{fR}}/{2}+\phi_2)}{|E_{f}E_{b}|+|E_{f}E_{bR}|\cos({\phi_{bR}}/{2}+\phi_1)+|E_{b}E_{fR}|\cos({-\phi_{fR}}/{2}+\phi_2)}
\label{eq4}
\end{equation}
\end{figure*}

\begin{figure*}[htbp]
\centering
\includegraphics[width=1\linewidth]{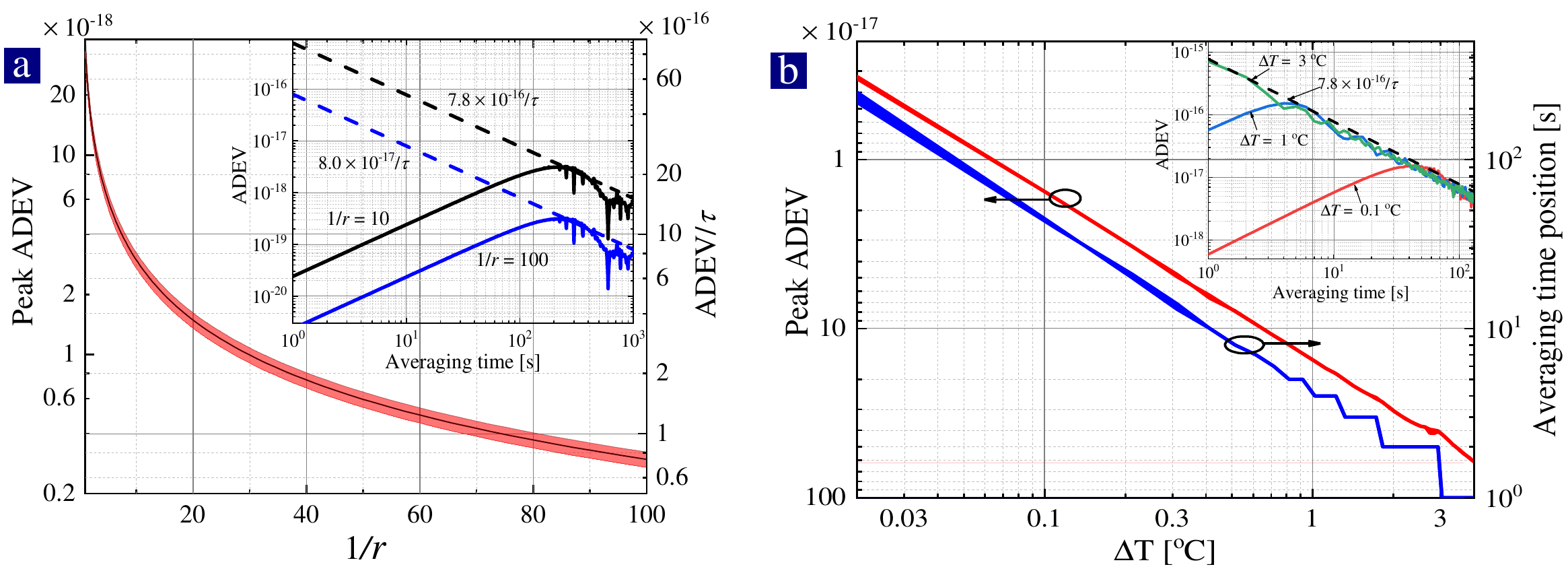}
\caption{ (a) Numerical simulation results of the peak  ADEV value (left axis) and the fitted $1/\tau$-dependence ADEV at the integration time of 1s (right axis) as a function of the inverse of the amplitude ratio denoted as $1/r=E_S/E_N$ at a fixed $\phi_N$ by setting $L=10$ m, $\Delta T=0.01$ $^{\circ}$C and $t_C=3,600$ s. The solid and dashed curves in the inset show the simulated ADEVs and fitted ADEVs with a dependence of $1/\tau$  at $r=0.1$ and $r=0.01$, respectively. (b)  Numerical simulation of the peak ADEV value ($\tau\geq1$ s, left axis) and the integration time position of the peak Allan deviation  (right axis) as a function of $\Delta T$ at a fixed $r=0.1$ with $L=10$ m, and $t_C=3,600$ s. The inset illustrates the ADEV at $\Delta T=0.1$ $^{\circ}$C (red curve), $\Delta T=1$ $^{\circ}$C (blue curve), and $\Delta T=3$ $^{\circ}$C (green curve), and the fitted ADEV (dashed curve) with a dependence of $7.8\times10^{-16}/\tau$.}
\label{fig3}
\end{figure*}

\subsection{Improved extraction setup}

Similar to the technique that was used for suppressing the backscattering induced phase noise by breaking the frequency symmetry on the main link, we propose two schemes that suppress the backscattering by locally breaking the frequency symmetry by inserting one AOM or two AOMs at the extraction point. In the first scheme as illustrated in Fig. \ref{fig2}(a), we replace the $2\times2$ optical coupler by an AOM denoted as AOM$_a$ with an angular frequency of $\omega_a$ and two $1\times2$ couplers at the extraction point, which has a capability to distinguish the desired signal ``\textcircled{\raisebox{-0.9pt}{3}}'' from other main noise terms ``\textcircled{\raisebox{-0.9pt}{1}}'',  ``\textcircled{\raisebox{-0.9pt}{2}}'' and ``\textcircled{\raisebox{-0.9pt}{4}}''. A main drawback related to this scheme is that inserting the AOM will modify the bidirectional optical frequency on the main fiber link, resulting in different beat-note at the local site for the servo controller. This issue can be solved by using the scheme illustrated in Fig. \ref{fig2}(b) where two AOMs are inserted at both sides of the $2\times2$ coupler with opposite sign of the AOM diffraction order and the same angular RF frequency of $\omega_a$. By setting this configuration, the net frequency coming from the extraction node will be null.

\section{Stability degradations due to backscattering}

\subsection{Studying the main limitation from the noise amplitude or phase}
In this section, we concentrate on the theoretical analysis of the effect of the backscattering on the transferring optical frequency signal \cite{burns1983rayleigh, sliwczynski2012bidirectional, lee2005bidirectional}.  Injecting a narrow linewidth laser results in the interference of the Rayleigh backscattered power from the distributed scatterers along the fiber. The resulting appearance of such a coherent interference trace is constant for a static fiber condition but changes locally as the temporal relation between scatterers as a function of applied temperature or strain changes  \cite{fleyer2015comprehensive, cahill2015superlinear, gysel1990statistical, staubli1991crosstalk}.  To observe these effects, here we assume that the phase noise $\phi_R(z,t)$ varies on a time scale that is significantly longer than the round-trip propagation time of the light in the fiber. Hence, the phase noise adds coherently during the backward and forward propagation and the two propagation phase terms  are approximately the same  \cite{fleyer2015comprehensive}, resulting in,
\begin{equation}
\begin{split}
{\varphi_R(z,t)}&\approx2\frac{\nu}{c}\left(L\frac{\partial n}{\partial T}+n\frac{\partial L}{\partial T}\right) T(t)\\
&=2\frac{L\nu}{c}\left(\alpha_n+n\alpha_{\Lambda}\right){ T(t)},
\end{split}
\label{eq5}
\end{equation} 
where $\nu$ is the carrier frequency,  $c$ is the speed of light, $\alpha_n\approx1.06\times10^{-5}$/K (room-temperature, 1550 nm) and $\alpha_\Lambda\approx 5.6\times10^{-7}$/K (room-temperature, 1550 nm) are, respectively, the thermo-optic and thermal expansion coefficient of the typical single-mode fiber, and $n$ is the effective refractive index of the fiber, $L$ is the fiber length and $T(t)$ is the temperature fluctuations. Here we assume $T(t)$ fluctuates as a sine function with the amplitude of $\Delta T$ and the cycle of $t_C$. Therefore, the Rayleigh backscattering phase fluctuation amplitude is proportional to the fiber length and temperature variation amplitude $\Delta T$. In a practical situation, the temperature fluctuations can be constructed by combining several sine functions with different periods and amplitudes \cite{pinkert2015effect}.

Before simulating the effect of the Rayleigh backscattering noise in a full fiber link, we first analyze the independent influence of the amplitude and the phase of the backscattering noise signal. Thus, we rewrite  Eq. \ref{eq3} into another form, which only consists of one signal term with the signal amplitude of $E_S$ and the null initial phase and one noise term with the amplitude $E_N$ and the phase $\phi_N$, resulting in 
\begin{equation}
{E_C}(t)\propto E_S\cos[\nu t]+E_N\cos[\nu t+\phi_N+\phi_o],
\label{eqxx}
\end{equation}
where the $\phi_o$ represents the initial phase offset between the signal  and  noise term.  We define the amplitude ratio as $r=E_N/E_S=\sqrt{P_N}/\sqrt{P_S}$ with ${P_S}$ and ${P_N}$ being the signal optical power and the noise optical power, respectively. we assume that the amplitude $E_N$ can be arbitrarily adjusted. Here the noise phase $\phi_N$ is calculated by using Eq. \ref{fig5} and we adjust it by changing the temperature amplitude. In this way, we can independently investigate the effect of the noise amplitude and phase on the system's performance.

To make the article consistent, we estimate and measure the $\Pi$-averaging fractional frequency instability in terms of Allan deviation (ADEV) in Sec.  III and in the next section (Sec. IV), respectively. Note that using the $\Pi$-averaging fractional frequency instability could have a  lower instability as discussed in \cite{dawkins2007considerations} and experimentally demonstrated in \cite{bercy2016ultrastable}.

\begin{figure}[hbtp]
\centering
\includegraphics[width=\linewidth]{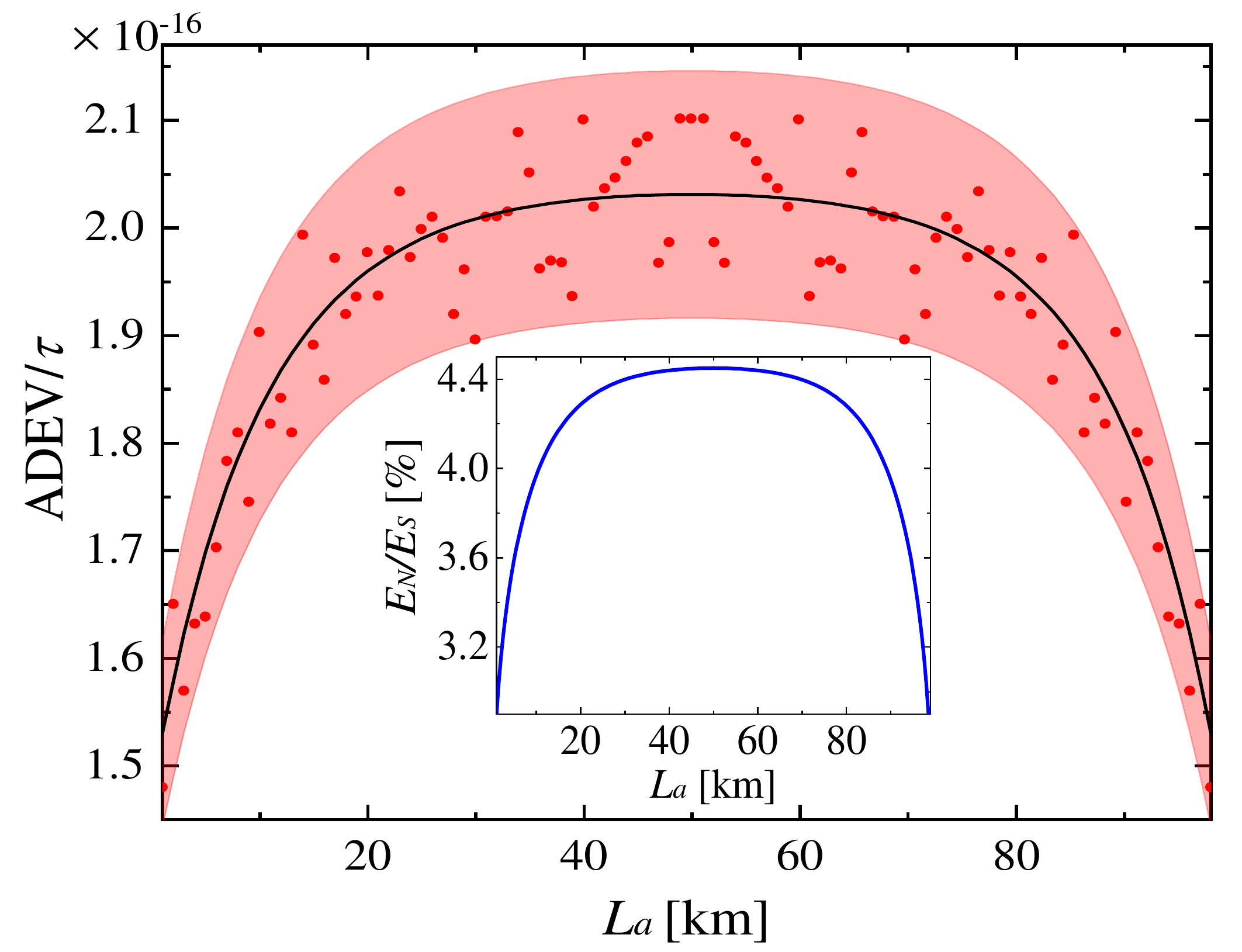}
\caption{Numerical simulation results of the ADEV  at the integration time of 1 s as a function of the fiber section length $L_a$ at $\Delta T =3$ $^{\circ}$C and $t_C=3,600$ s. The uncertainty illustrated by the shaded area comes from the initial phase offsets $\phi_1$ and $\phi_2$ as depicted in Eq. \ref{eq4} and the scatters represent the initial phase offsets of $\phi_1=\phi_2=0$. We can see that the access node close to the local and remote site performs a lower ADEV of $\sim1.6\times10^{-16}/\tau$ and increases up the $\sim2.2\times10^{-16}/\tau$ at $L_a=50$ km. The inset shows the ratio of the noise terms and the signal term as a function of $L_a$. $E_N$ ($|E_fE_{bR}|+|E_bE_{fR}|$) and $E_S$ ($|E_fE_b|$) are the total electrical field of the noise terms and the desired term  shown in Eq. \ref{eq3}.}
\label{fig4}
\end{figure}

Figure \ref{fig3}(a) and (b) illustrate the peak value of  the ADEV as functions of the amplitude ratio $r$ and the temperature fluctuation amplitude  $\Delta T$. The shaded areas are taking into account the initial phase offset $\phi_0$. In the first simulation as shown in Fig. \ref{fig3}(a), we set a constant  $\phi_N$ by choosing $\Delta T =0.01$ $^{\circ}$C, $t_C = 3,600 $ s and $L=10$ m. In this configuration, we have the capability to observe the evolution in the peak value of the ADEV for different parameters and to examine the main limitation of the backscattering as shown the insets of Fig. \ref{fig3}. We can clearly see in Fig. \ref{fig3} (a) for a fixed $\phi_{N}$ the peak values of the ADEV are located at around 250 s and the optical frequency transfer performance is significantly affected by the amplitude ratio $r$, demonstrating that the amplitude ratio $r$ has a certain effect on the ADEV. When $r=0.03$ corresponding to the noise optical power $10\text{log}_{10}(P_S/P_N)=30$ dB lower than the signal optical power, the ADEV is approximately $\sim 2.4 \times 10^{-16}$, which is better than even the state-of-the-art multiple-access optical frequency transfer systems \cite{bercy2014line, grosche2014eavesdropping, bercy2016ultrastable}.  Figure \ref{fig3}(b) illustrates the calculated peak value of the ADEV $(\tau\geq1$ s) with different temperature variations $\Delta T$ at $ t_C =3,600$ s, $L=10$ m and $r=0.1$. We note that with the increase of $\Delta T$ leads to the increase of the amplitude of $\phi_{N}$ and the peak value of the ADEV shifts to the short-term as the blue curve illustrated in Fig. \ref{fig3}(b). In this case, the short-term stability is growing gradually whereas the long-term stability is scarcely changed as depicted in the inset of Fig. \ref{fig3}(b).  Considering factors of $r$ and $\phi_{N}$, it is important to stress that the isolation requirement for lossless frequency transfer is mainly determined by $r$ rather than $\phi_{N}$ since $r$ mainly affects the amplitude of the $1/\tau$ dependence ADEV while $\phi_{N}$ only affects its position. Consequently, to avoid the backscattering effect, it is a priority to reduce the noise amplitude with respect to the signal amplitude instead of minimizing the phase noise.

\subsection{100 km full link simulation}


\begin{figure*}[htbp]
\centering
\includegraphics[width=1\linewidth]{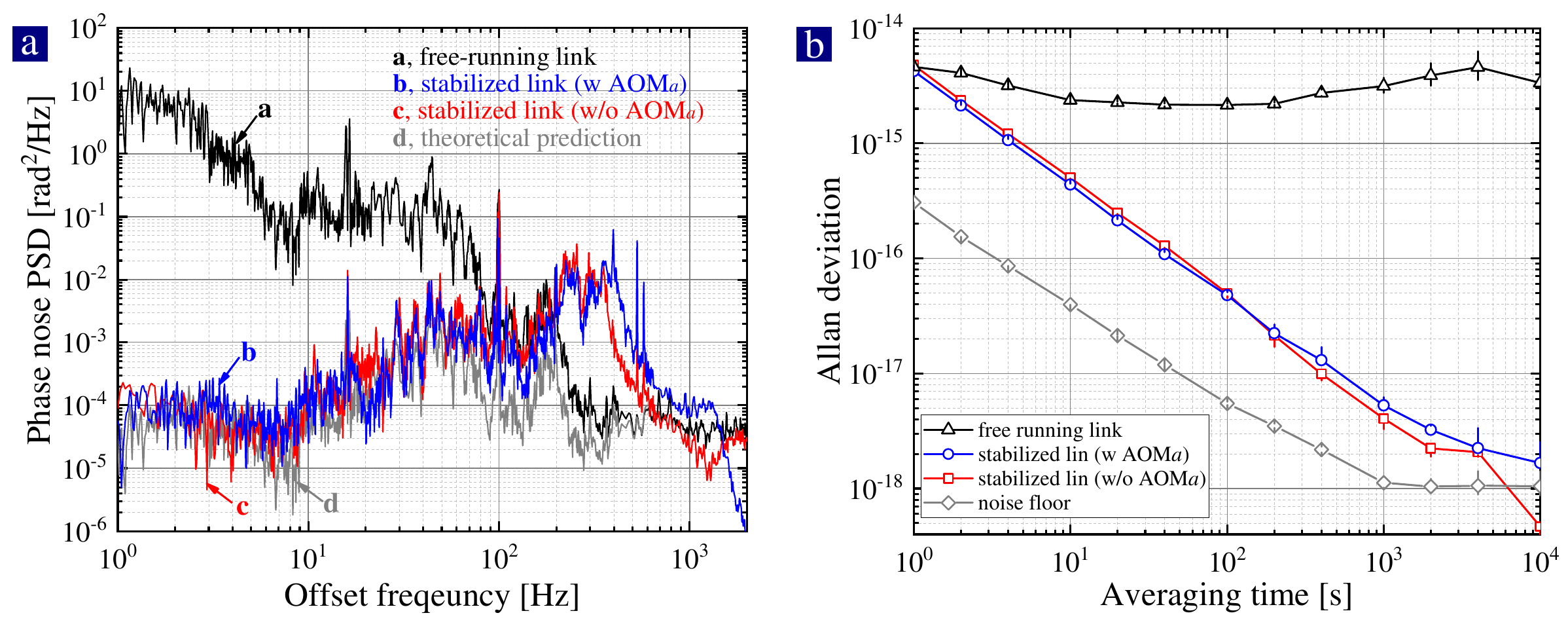}
\caption{(a) Measured main fiber link phase noise PSD for a 100 km free-running (\textbf{a}, black curve) and stabilized link with AOM$_a$ (\textbf{b}, blue  curve) and without  AOM$_a$ (\textbf{c}, red  curve). The gray dashed curve (\textbf{d}) is the theoretical prediction. (b) The corresponding main fiber link ADEV taken with a $\Pi$-type counter for a 100 km free-running (black triangle markers) and stabilized link with AOM$_a$ (blue circle markers) and without AOM$_a$ (red square markers). The gray diamond markers illustrate the system noise floor. }
\label{fig5}
\end{figure*}

To simulate a more practical case with the aim of comparing the behavior of the Rayleigh backscattering to the experimental results performed in the next section, we calculate the Rayleigh backscattering electrical field as,
\begin{equation} 
\begin{split}
E_R(t)=&\int_0^{L}E_0\left(t-\frac{2nz}{c}\right)\kappa(z)\varrho(z,t)e^{-2\alpha z}e^{j\varphi_R(z,t)}dz,
\end{split}
\label{fig17}
\end{equation}
where $\alpha$ is the intensity attenuation coefficient, $\kappa(z)$ is the backscattering coefficient, $E_0(t)$ denotes the electrical field of the laser source, $\varphi_R(z,t)$ is the total phase fluctuations induced by the fiber during the forward and backward propagation as depicted in Eq. \ref{eq5} and $\varrho(z,t)$ is used to account for the influence of polarization.  In a practical case, the Rayleigh backscattering light is known to be partially depolarized \cite{van1993polarization}. In addition, randomly varying birefringence of field-deployed optical fibers results in some scrambling of the polarization of propagating signals \cite{karlsson2000long, schiano2004pmd}. To maximize the effect of the Rayleigh backscattering, we assume the same polarizations ($\varrho(z,t)=1$) of interfering fields between the local and backscattering lights.

 To avoid nonlinear effects, the signal power launched into the fiber link is kept low (6 mW in our case). Each optical coupler and each AOM have 3 dB and 2 dB losses, respectively. $E_{bR}$ and $E_{fR}$ shown in Eq. \ref{eq3} are calculated by using Eq. \ref{fig17}, and $E_b$ and $E_f$ are estimated by taking the fiber loss of $\alpha=0.2$ dB/km into account. The double Rayleigh scattering has been ignored in the simulation. Figure \ref{fig4} illustrates the ADEV with different fiber section length $L_a$ over a 100 km fiber link at $\Delta T=3$ $^{\circ}$C.  The noise and signal terms can interfere constructively or destructively depending on the initial phase offsets $\phi_1$ and $\phi_2$ in Eq. \ref{eq1}, making the phase  compensated signal to vary consequently. The scatters shown in Fig. \ref{fig4} represent typical initial phase offsets of $\phi_1=\phi_2=0$. With the continuous increase of the fiber length $L_a$ from 1 km to 50 km, the proportion of the Rayleigh scattering induced  electrical signal  gradually increases as shown in the inset of the Fig. \ref{fig4}, resulting in  the worse ADEV. This result is consistent with the one shown in Fig. \ref{fig3}(a). Note that the fractional frequency stability of the second half of the fiber section $L_b$ will be decreased from 50 km and reaches the minimum at 100 km. It can also be seen that when the fiber length $L_a$ ($\leq50$ km) increases to a certain degree, the ADEV at the integration time of 1 s changes slowly. Consequently, when the fiber length $L_a$ is longer than 30 km, the ADEV reaches a steady state and a stability of better than $\sim2.2\times10^{-16}/\tau$ even if for a $L=100$ km fiber link. This result is consistent with the simulation results shown in Fig. \ref{fig3}(a) at the $r\simeq0.044$. By taking the polarization fluctuations into account, the value will be even better. However, this stability is much lower than even the sate-of-the-art multiple-access optical frequency transfer systems \cite{bercy2014line, grosche2014eavesdropping, bercy2016ultrastable}. Therefore, we can conclude that the backscattering noise has limited effect on the stability of optical frequency dissemination.  

\section{Experimental apparatus and results}
\subsection{Experimental apparatus}

To verify the effect of the Rayleigh backscattering on the fiber-optic optical frequency transfer performance, we set up optical frequency transfer systems with and without the backscattering, which were achieved by whether to insert an AOM (AOM$_a$) at the extraction point as illustrated in Fig. \ref{fig2} (a). The proposed scheme was tested by using a narrow-linewidth optical source (NKT X15) at a frequency near 193 THz with a typical linewidth of 100 Hz. The signal was transmitted to a remote site along a 100 km spooled fiber link consisting of two 50 km fiber sections ($L_a$ and $L_b$). To boost the fading optical signal after the 100 km fiber link, we install one home-made bidirectional erbium-doped fiber amplifier (EDFA) at the remote site. At the same time, we install one more EDFA after 50 km for just compensating the insertion losses coming from the additional AOM$_a$ and the optical coupler. In the view of the laboratory conditions, we choose the scheme illustrated in Fig. \ref{fig2}(a) to set up the experimental system. To better measure the effect of the backscattering on the fiber-optic frequency transfer performance, we select the point where ADEV deteriorated the most according to the simulation results in Sec. III, that is $L_a=50$ km. Here we set AOM$_1=80$ MHz (downshifted mode, $-1$ order), AOM$_2=90$ MHz (upshifted mode, $+1$ order), AOM$_3=40$ MHz (upshifted mode, $+1$ order) and AOM$_a=40$ MHz (downshifted mode, $-1$ order). With this configuration, we can acquire beat-note {$2\omega_r-\omega_a=140$ MHz at the access node (Fig. \ref{fig2}(a)) and $2\omega_r=180$ MHz at the access node (Fig. \ref{fig1}) for the main fiber link with AOM$_a$ and without AOM$_a$, respectively.  We use a frequency counter, which is referenced to the RF frequency source at the local site, to record the beating frequencies between the fiber input light and the output light at the remote site and the access node, respectively. Additionally, to measure the phase noise of the optical carrier frequencies at both remote site and access node, we perform the measurement by feeding the heterodyne beat frequency together with a stable frequency reference to a phase detector. The voltage fluctuations at the phase detector output are then measured with a fast Fourier transform (FFT) analyzer to obtain the phase fluctuations $S_{\phi}(\omega)$. 



\subsection{Experimental results-Main fiber link}
Figure \ref{fig5}(a) shows the measured  phase noise PSD for the free-running fiber link (\textbf{a}, black curve) and for the stabilized link where phase noise cancellation was active for both cases with (\textbf{b}, blue curve) and without AOM$_{a}$ (\textbf{c}, red curve), respectively.  We find that unlocked phase noise on our fiber link for both cases approximately follows a power-law dependence, $S_{\phi}(f)\simeq h_0/f^2$ rad$^2$/Hz with $h_0\simeq 40$ in our fiber link, for $f<1$ kHz, demonstrating that the white frequency noise is dominating in the free-running fiber link. By activating the fiber noise cancellation system, the phase noise PSD is largely suppressed down to $\sim5\times10^{-4}$ rad$^2$/Hz within its delay-unsuppressed bandwidth $\sim 300$ Hz, illustrating the remaining noise is mainly determined by the white phase noise for both cases. Note that the shifted bump position from the ideal position of $1/4\tau_0\simeq500$ Hz is mainly due to the time delay of electrical components and unoptimized parameters of the servo controller. From Fig. \ref{fig5}(a), we can also find the measured locked phase noise to be in good agreement with the theoretical value (\textbf{d}, gray curve) as predicted by the formula in \cite{bercy2014line}.

\begin{figure*}[htbp]
\centering
\includegraphics[width=1\linewidth]{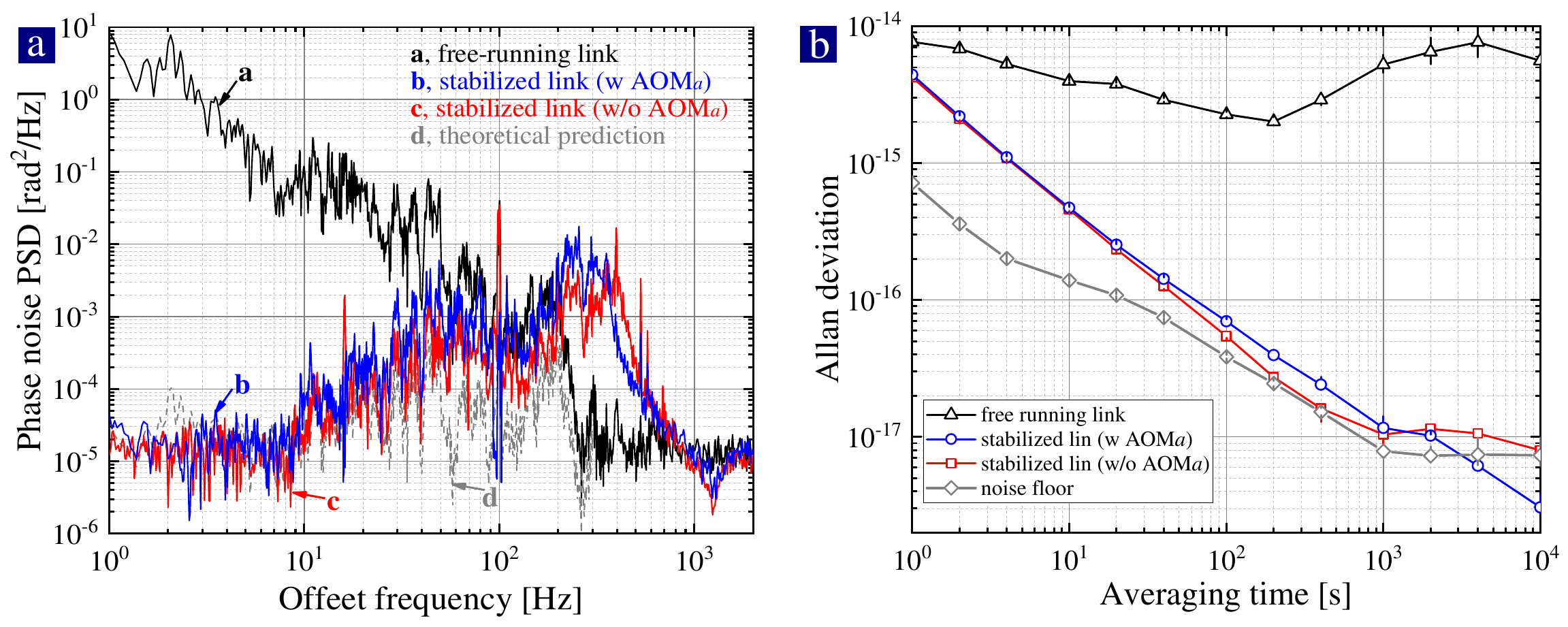}
\caption{(a) Measured access node ($L_a/L_b=50/50$ km) phase noise PSD for a 100 km free-running (\textbf{a}, black curve) and stabilized link with AOM$_a$ (\textbf{b}, blue  curve) and without  AOM$_a$ (\textbf{c}, red  curve). The gray dashed curve (\textbf{d}) is the theoretical prediction. (b) The corresponding measured access node ($L_a/L_b=50/50$ km)  ADEV countered with a non-averaging $\Pi$-type counter for a 100 km free-running (black triangle markers) and stabilized link with AOM$_a$ (blue circle markers) and without AOM$_a$ (red square markers). The gray diamond markers illustrate the system noise floor.}
\label{fig7}
\end{figure*}

\begin{figure}[htbp]
\centering
\includegraphics[width=\linewidth]{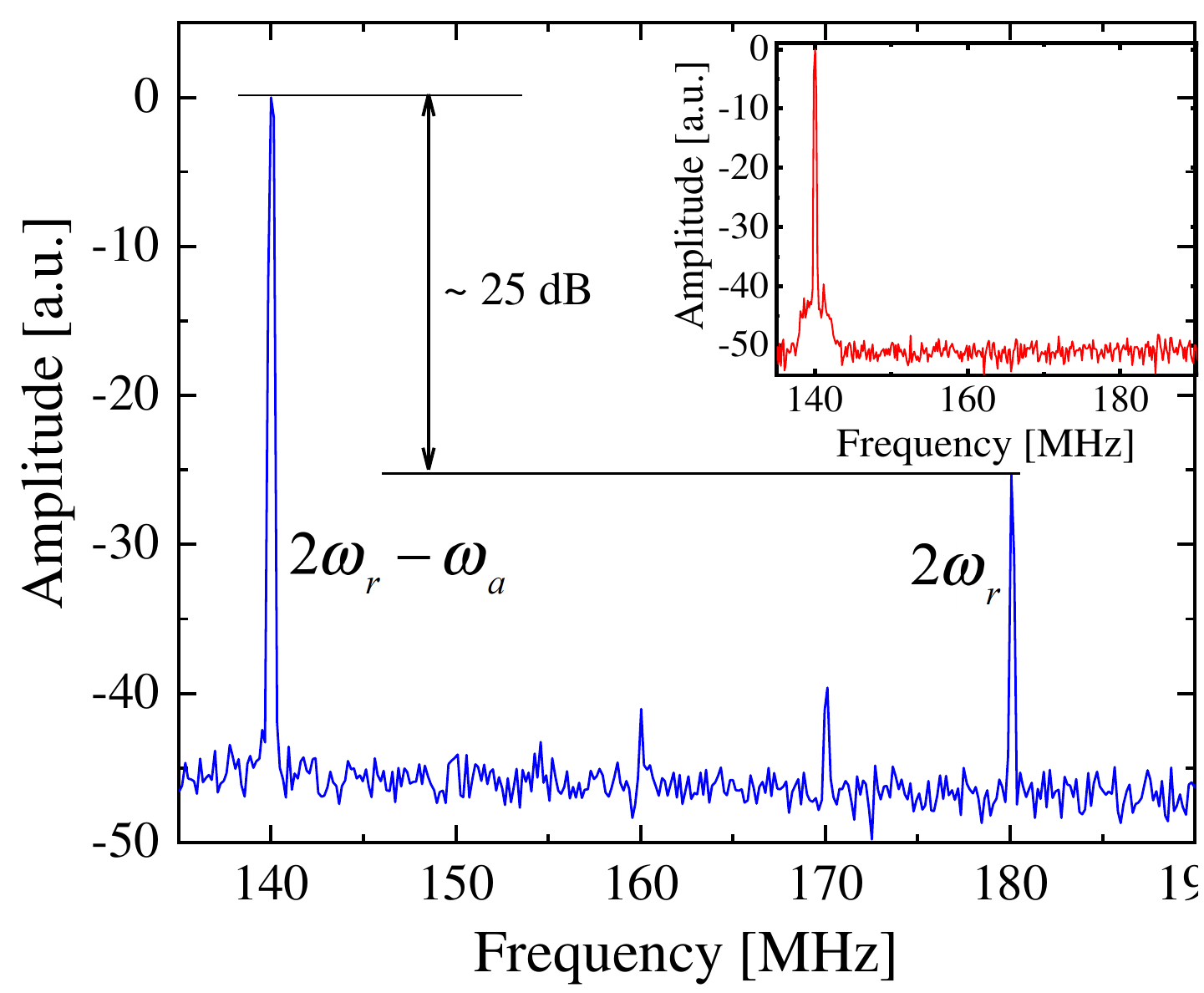}
\caption{Typical power spectrum of  the  interference of  $E_{b}(t)$ with $E_f(t)$ ($2\omega_r-\omega_a=140$ MHz) and $E_{b}(t)$ with $E_{fR}(t)$  ($2\omega_r=180$ MHz)  after the PD2 measured by power spectrum analyzer with a resolution bandwidth (RBW) of 100 kHz. Other spikes could be coming from multiple reflections in the fiber and nonlinear effect of the photo-detector and amplifiers. The inset shows the power spectrum after an RF bandpass filter with a 3 dB bandwidth of 5 MHz. The frequency centered on 140 MHz can be effectively filtered out by the bandpass filter. We experimentally observed that the noise terms have great amplitude fluctuations over 10 dB. Other noise terms between  the extracted optical signals with the backscattered noise signals have similar results.}
\label{fig6}
\end{figure}

Complementary to the frequency-domain feature, a time-domain characterization of the fractional frequency instability is shown in Fig. \ref{fig5} (b).  The fractional frequency instability has been acquired for the stabilized fiber link with (blue circle markers) and without (red square markers) AOM$_a$ calculated by the ADEV taken with the counters operating in a nonaveraging $\Pi$-type with a 1 s gate time \cite{dawkins2007considerations}.   For both cases with and without AOM$_a$, the fractional frequency instability in terms of the ADEV is around $4.4\times10^{-15}$ at an averaging time of 1 s and reaches a floor in the $10^{-18}$-range for a few thousand of seconds of averaging time. In general, the fractional frequency instability of the two cases is similar except for a long period of averaging time that we attribute the slight difference for both cases to differential temperature fluctuations in the local and  the remote optical setup measured by different times, indicating that whether the AOM is inserted at the extraction point does not have much impact on the main fiber link.

\subsection{Experimental results-Access node}


To examine the backscattering induced RF spectrum, we first measured the power spectrum $E_{\text{RF}}$ detected by PD2 at the access node with AOM$_a$ before and after a bandpass filter centered on 140 MHz as shown in Fig. \ref{fig6}. For clarity, we select one of the noise terms  to display. The noise signal produced by mixing $E_b(t)$ and $E_{fR}(t)$ as ``\textcircled{\raisebox{-0.9pt}{1}}'' illustrated in Fig. \ref{fig2}(a) coming from  the Rayleigh backscattering, the fresnel reflection and the amplified spontaneous emission (ASE) noise of the EDFAs is only -25 dB lower than the desired signal in our typical experimental configuration. We can clearly see that the bandpass filter can suppress all unwanted signals by at least 50 dB below the signal of interest as the inset of Fig. \ref{fig6}. The unwanted signals could be further rejected in the frequency dividing process and by a power amplifier at the output of the bandpass filter.  We experimentally observed that the noise terms have great amplitude fluctuations over 10 dB. Other noise terms between the extracted optical signals with the backscattered noise signals have similar results. Therefore, in our experiment, we compare two configurations: one without AOM$_a$ as illustrated in Fig. \ref{fig1} and the other with AOM$_a$ as shown in Fig. \ref{fig2}(a).

Figure \ref{fig7}(a) shows the phase noise PSD for the out-of-loop beat signal of the unstabilized and stabilized interferometer at access node with (\textbf{b}, blue curve) and without AOM$_a$ (\textbf{c}, red curve), respectively. Similar to the main fiber link, the phase noise of the unstabilized interferometer (\textbf{a}, black curve)  at the access node  is dominated by $S_{\phi}(f)\simeq h_0/f^2$ rad$^2$/Hz up to a Fourier frequency of $f\simeq1$ kHz while above 1 kHz a white phase noise level of  $S_{\phi}(f)\simeq3\times10^{-5}$ rad$^2$/Hz is obtained. In comparison with the main fiber link, the coefficient $h_0$ is proportional to the fiber link length and is approximately 20 for the access node. Once the main fiber link is stabilized, the fluctuations of the path length and those arising from paths that are inside of the loop will be corrected automatically at the access node. The phase noise is suppressed more than 50 dB at 1 Hz. Additionally, we observed the similar performance for both with and without AOM$_a$ in terms of phase noise PSDs. 

A time-domain characterization of frequency stability (ADEV) is shown in Fig. \ref{fig7}(b). In this plot, curves with the circle and square markers indicate the fractional frequency stability of stabilized optical carrier frequency dissemination at access node with (blue circle markers) and without (red square markers) AOM$_a$, respectively. With the implementation of fiber noise cancellation at the main link and the access node, both optical frequency transfer configurations achieve a fractional frequency stability of $\sim4.2\times10^{-15}$ at the integration of $1$ s and $\sim1.1\times10^{-17}$ at $1,000$ s. We can clearly see that the results obtained under the two setups are similar, whether or not we actively suppress the backscattered noise. However, it also shows that the system performs one order of magnitude worse than the simulation results in Sec. III, which should be about  $2.2\times10^{-16}/\tau$. To experimentally observe the backscattering effect, one way is to select a quieter  fiber, resulting in a lower residual fiber phase noise. This seems impossible in particular for field-deployed fiber links \cite{newbury2007coherent, grosche2009optical, foreman2007coherent, jiang2008long, yamaguchi2011direct, calosso2014frequency}. Another way is to boost the noise term as discussed in Sec. III. Unfortunately, once we introduce strong phase noise by injecting part of uncompensated light, the system components like the frequency divider  start to stop working.


In comparison with the main fiber link shown in Fig. \ref{fig5}, we observed a similar short-term stability and a worse long-term stability at the access node. We attribute this to more optical components adopted at the access node and it is difficult to match the length of the fiber paths outside the loop. To confirm this, we measured, respectively, the noise floor of the main link (Fig. \ref{fig5}(b)) and the access node (Fig. \ref{fig7}(b)) by replacing the fiber spool with a 1 m fiber plus corresponding attenuators. We can observe that the floor of the main link (the access node) with a stability of $\sim3.1\times10^{-16}$ ($7.1\times10^{-16}$) at 1 s and $\sim1.1\times10^{-18}$ ($7.9\times10^{-18}$) at 1,000 s is obtained. Consequently, the ADEV deterioration of the access node compared with the main link is indeed caused by the unmatched optical paths \cite{foreman2007coherent, predehl2012920, droste2013optical2}.  In the current setup, the interferometer was built with an additional fiber-pigtailed AOM (AOM$_3$) and other optical couplers, which caused non-optimal spatial design and thus involved relatively long uncompensated fibers and thermal effect. In the future, we are also planning to stabilize the uncompensated path with passive or active methods or with free-space components \cite{predehl2012920, droste2013optical2, droste2014optical}. 


 
 
 \section{Discussion}



Although breaking the frequency symmetry at the access point can be  used to avoid the backscattering at the access nodes, the frequency difference between the forward and backward propagation could not be arbitrarily  large. The frequency difference between forward and backward propagation $\nu_s$ will lead to unequal phase noise fluctuations $\phi_p$ on the bidirectional signals. Assuming the link noise introduces an equally delay fluctuation $\Delta\tau$ for both forward and backward signals, the phase fluctuations of the backward propagation signal is,
\begin{equation}
\Delta\phi_b=\Delta\tau\times\nu_b=\Delta\tau\times\nu_f(1+\frac{\nu_s}{\nu_f})=\phi_f(1+\frac{\nu_s}{\nu_f})
\end{equation}
where $\phi_f$ and $\phi_b$ are forward and backward phase fluctuations, respectively, and $\nu_f$ and $\nu_b$ are frequencies of the forward and backward propagation signal. Consequently, the residual uncorrected phase fluctuation $\phi_f+\phi_c$ results in the instability of the compensated link as $\sigma_y(\tau)\approx a \sigma_y(\tau)_{free}$ with $\sigma_y(\tau)_{free}$ being the free-running link instability. In our configuration $\nu_s=180$ MHz, this leads to $\sigma_y(\tau)\approx 5\times10^{-7} \sigma_y(\tau)_{free}$. Therefore, we have to notice the frequency difference between two directions when multiple access nodes are installed along the fiber link with the scheme as shown in Fig. \ref{fig2}(a)


A common figure of merit for a long-distance optical frequency transfer system is its long term stability in terms of the ADEV. In this work, we have ignored other noises including laser phase and amplitude noise, the amplitude noise induced by the backscattering signal. These noises are typically lower than the residual delay-limited fiber phase noise \cite{eliyahu2008rf, cahill2017additive, williams2008high}. For a relatively short distance optical frequency system such as optical frequency comparison with optical frequency combs, two limiting regions of phase/frequency noise are interesting: low Fourier frequency with a large phase range (slow frequency noise) and high Fourier frequency with a small phase range (fast frequency noise).  Currently, much attention has been paid to the relative frequency instability and accuracy, namely slow frequency noise and little has been paid to the line-width of the transferred light. Transfer of coherent laser light to remote sites while maintaining its high spectral purity (fast and slow frequency noise) and low phase noise with passive phase noise correction \cite{hu2020multi} is permitting a variety of applications such as high-resolution spectroscopy, precision detection and measurements \cite{foreman2007remote} and enables distribution and correlation of spatially and spectrally separated coherent signals with the assistance of optical frequency combs \cite{coddington2007coherent, xie2017photonic, diddams2007molecular, argence2015quantum}.  The effect on the short-term stability, determining the spectral purity transfer limitation imposed by this backscattering induced amplitude and phase noise could be addressed through a combination of modeling and measurements for amplitude and phase noise generated in the loop \cite{fleyer2015comprehensive, eliyahu2008rf, cahill2017additive}. Such experiments would be an interesting topic for further study.

\section{Conclusion}

In summary, we demonstrated a fundamental concept that the Rayleigh backscattering from disorder within the optical fiber can be effectively suppressed by locally breaking frequency symmetry. We also proposed a method for estimating interfering signals, which appear at the access nodes over a bidirectional fiber link because of the Rayleigh backscattering effect on the fiber-optic optical frequency transfer system. The numerical simulation results indicate that the degradation of the $1/\tau$-dependence ADEV is mainly  dependent on the amplitude ratio between the Rayleigh backscattering and the desired signal, and the peak position $\tau$ goes inversely with the amplitude of the temperature fluctuations and fiber link length. We have performed an experiment on a 100-km multiple-access 193 THz optical frequency transfer system. The measured results are given to confirm that the effect of the crosstalk can be neglected at the precision of our setup, illustrating that it is possible to realize a multi-access optical frequency transfer system with commercially available connectors in which the Rayleigh backscattering is the dominant crosstalk contribution. Multiple-access optical frequency dissemination opens a way to a wide distribution of an ultrastable frequency reference, enabling applications beyond metrology, and the development of new high-sensitivity experiments in a broad range of applications.


\ifCLASSOPTIONcaptionsoff
  \newpage
\fi



%
\bibliographystyle{IEEEtran}
\bibliography{Optics}
\begin{IEEEbiographynophoto}{Liang Hu}
received the B.S. degree from Hangzhou Dianzi University, China, in 2011, and the M.S. degree from Shanghai Jiao Tong University, China, in 2014. He received the Ph.D. degree from University of Florence, Italy, in 2017 during which he was a Marie-Curie Early Stage Researcher at FACT project. He is currently a Tenure-Track Assistant Professor in the State Key Laboratory of Advanced Optical Communication Systems and Networks, Department of Electronic Engineering, Shanghai Jiao Tong University, China. His current research interests include photonic signal transmission and atom interferometry.
\end{IEEEbiographynophoto}

\begin{IEEEbiographynophoto}{Xueyang Tian}
received the B.S. degree from Shanghai Dianji University, China, in 2017. She is currently a graduate student in the State Key Laboratory of Advanced Optical Communication Systems and Networks, Department of Electronic Engineering, Shanghai Jiao Tong University, China. Her current research interests include photonic signal transmission.
\end{IEEEbiographynophoto}

\begin{IEEEbiographynophoto}{Guiling Wu}
received the B.S. degree from Haer Bing Institute of Technology, China, in 1995, and the M.S. and Ph.D. degrees from Huazhong University of Science and Technology, China, in 1998 and 2001, respectively. He is currently a Professor in the State Key Laboratory of Advanced Optical Communication Systems and Networks, Department of Electronic Engineering, Shanghai Jiao Tong University, China. His current research interests include photonic signal processing and transmission.
\end{IEEEbiographynophoto}


\begin{IEEEbiographynophoto}{Jianguo Shen}
received his Bachelor Degree on Physics Educations from Zhengjiang Normal University in 2002, Master Degree on Circuit and system from Hangzhou Dianzi University in 2007, Ph.D on Electromagnetic and Microwave Technology from Shanghai Jiaotong University in 2015. Currently, he is an associate professor at Zhejiang normal university of china. His research interests include microwave photonic signals processing and time and frequency transfer over the optical fiber.   
\end{IEEEbiographynophoto}

\begin{IEEEbiographynophoto}{Jianping Chen}
received the B.S. degree from Zhejiang University, China, in 1983, and the M.S. and Ph.D. degrees from Shanghai Jiao Tong University, China, in 1986 and 1992, respectively. He is currently a Professor in the State Key Laboratory of Advanced Optical Communication Systems and Networks, Department of Electronic Engineering, Shanghai Jiao Tong University. His main research interests include opto-electronic devices and integration, photonic signal processing, and system applications. He is a Principal Scientist of National Basic Research Program of China (also known as 973 Program).
\end{IEEEbiographynophoto}

\end{document}